\documentclass[conference]{IEEEtran}
\def\ps@headings{%
\def\@oddhead{\mbox{}\scriptsize\rightmark \hfil \thepage}%
\def\@evenhead{\scriptsize\thepage \hfil \leftmark\mbox{}}%
\def\@oddfoot{}%
\def\@evenfoot{}}
\makeatother
\usepackage{booktabs} % For formal tables
\usepackage{url}
\usepackage{svg}
\usepackage{etoolbox}
\usepackage{cuted}
\usepackage{lipsum}
\usepackage{amssymb}
\usepackage{array}
\usepackage{amsmath}
\usepackage{tikz}
\usepackage{color}
\usepackage{dirtytalk}
\usepackage{algpseudocode}
\usepackage{algorithm}
\usepackage{graphicx}
\usepackage{caption}
\usepackage{mdframed}
\usepackage{tcolorbox}
\usepackage{comment}

\IEEEoverridecommandlockouts
\IEEEpubid{\makebox[\columnwidth]{978-1-7281-1328-9/19/\$31.00~\copyright2019 IEEE \hfill} \hspace{\columnsep}\makebox[\columnwidth]{ }}

\ifCLASSINFOpdf
\else
\fi
\DeclareRobustCommand*{\IEEEauthorrefmark}[1]{\raisebox{0pt}[0pt][0pt]{\textsuperscript{\footnotesize #1}}}

\begin{document}
%
% paper title
% Titles are generally capitalized except for words such as a, an, and, as,
% at, but, by, for, in, nor, of, on, or, the, to and up, which are usually
% not capitalized unless they are the first or last word of the title.
% Linebreaks \\ can be used within to get better formatting as desired.
% Do not put math or special symbols in the title.
% \title{Blockchain-based Access Control for Electronic-Health Records with Edge Computing}
\title{Attribute-based Multi-Signature and Encryption for EHR Management: A Blockchain-based Solution}

% author names and affiliations
% use a multiple column layout for up to three different
% affiliations

\author{\IEEEauthorblockN{Hao Guo\IEEEauthorrefmark{1}~~~~Wanxin Li\IEEEauthorrefmark{2}~~~~Ehsan Meamari\IEEEauthorrefmark{1}~~~~Chien-Chung Shen\IEEEauthorrefmark{1}~~~~Mark Nejad\IEEEauthorrefmark{2}}
\IEEEauthorblockA{\IEEEauthorrefmark{1}Department of Computer and Information Sciences~~~~~\IEEEauthorrefmark{2}Department of Civil and Environmental Engineering\\
University of Delaware, U.S.A. \\
\
\{haoguo,wanxinli,ehsan,nejad,cshen\}@udel.edu}}

% conference papers do not typically use \thanks and this command
% is locked out in conference mode. If really needed, such as for
% the acknowledgment of grants, issue a \IEEEoverridecommandlockouts
% after \documentclass

% for over three affiliations, or if they all won't fit within the width
% of the page, use this alternative format:
% 
%\author{\IEEEauthorblockN{Michael Shell\IEEEauthorrefmark{1},
%Homer Simpson\IEEEauthorrefmark{2},
%James Kirk\IEEEauthorrefmark{3}, 
%Montgomery Scott\IEEEauthorrefmark{3} and
%Eldon Tyrell\IEEEauthorrefmark{4}}
%\IEEEauthorblockA{\IEEEauthorrefmark{1}School of Electrical and Computer Engineering\\
%Georgia Institute of Technology,
%Atlanta, Georgia 30332--0250\\ Email: see http://www.michaelshell.org/contact.html}
%\IEEEauthorblockA{\IEEEauthorrefmark{2}Twentieth Century Fox, Springfield, USA\\
%Email: homer@thesimpsons.com}
%\IEEEauthorblockA{\IEEEauthorrefmark{3}Starfleet Academy, San Francisco, California 96678-2391\\
%Telephone: (800) 555--1212, Fax: (888) 555--1212}
%\IEEEauthorblockA{\IEEEauthorrefmark{4}Tyrell Inc., 123 Replicant Street, Los Angeles, California 90210--4321}}

% use for special paper notices
%\IEEEspecialpapernotice{(Invited Paper)}

% make the title area
\maketitle

% As a general rule, do not put math, special symbols or citations
%\section{% in the abstract}
\begin{abstract}
%Nowadays, attribute-based encryption (ABE) and attribute-based signature (ABS) are regarded as  critical technology to address data security and user authenticity. However,in all traditional ABE and ABS mechanism, the centralized private key generator (PKG) has the ability to generate private keys, which can bring risks such as key abuse and sensitive data leakage. Also, the traditional cloud storage platform usually executes in a centralized storage manner, so a single point of failure may lead to the fail of the storage system. With the inherent decentralized characteristics of blockchain technology, we transfer the traditional centralized ABE and ABS system into decentralized ABE and attribute-based multi-signature (ABMS) scheme by designing a multi-authority authentication and authorization scheme.

%In this paper, we propose the EHR data storage and sharing mechanism and design a framework which integrates the edge storage platform, the Hyperledgerblockchain network, and ABMS technology.

%In this framework, the data owner has the ability to encrypt EHR data by specifying access policy using his/ her special attributes which is issued by different attribute authorities, and achieve the data owner authenticity property without leaking the sensitive identity information.  

The global Electronic Health Record (EHR) market
is growing dramatically and has already hit \$31.5 billion
in 2018. To safeguard the security of EHR data and privacy %issues 
of patients, fine-grained information access and sharing mechanisms are essential for EHR management. This paper proposes a hybrid architecture of blockchain and edge nodes to facilitate EHR management.
In this architecture, we utilize attribute-based multi-signature (ABMS) scheme to authenticate user's signatures without revealing the sensitive information and multi-authority attribute-based encryption (ABE) scheme to encrypt EHR data which is stored on the edge node.
%a blockchain-based controller authenticate patients by a multi-authority attribute-based multi-signature (ABMS) scheme and 
%Blockchain serves as a tamper-proof log to record authentication events and access activities into transactions.
%In addition, off-chain edge nodes store the EHR data which is encrypted by a multi-authority attribute-based encryption (ABE) scheme.
We develop the blockchain module on Hyperledger Fabric platform and the ABMS module on Hyperledger Ursa library. We measure the signing and verifying time of the ABMS scheme under different settings, and experiment with the authentication events and access activities which are logged as transactions in blockchain.
% and the performance of blockchain transaction processing time under both author
%The results of the extensive experiments show that the system performance can meet the requirements (e.g., response time) of real-world applications while protecting EHR data security, and it is robust against unauthorized retrievals. 

\end{abstract}


% no keywords

% For peer review papers, you can put extra information on the cover
% page as needed:
% \ifCLASSOPTIONpeerreview
% \begin{center} \bfseries EDICS Category: 3-BBND \end{center}
% \fi
%
% For peerreview papers, this IEEEtran command inserts a page break and
% creates the second title. It will be ignored for other modes.
\IEEEpeerreviewmaketitle

\IEEEpubidadjcol

\section{Introduction}

Electronic health records (EHR), although containing private information for patient diagnosis and treatment, need to be frequently shared among different participants, such as healthcare providers, insurance companies, pharmacies, and medical researchers~\cite{dubovitskaya2017secure}. Therefore, one big challenge to EHR management is to gather, store and share personal healthcare information without violating privacy or compromising security.
%To address this challenge, the US Health Insurance Portability and Accountability Act (HIPAA), for instance, sets guidelines to modernize and protect the flow of healthcare information. 
In the context of EHR data management, although healthcare providers (termed data users), such as doctors, nurses, phlebotomists, medical laboratory scientists, pathologists, etc., need to authenticate the patients (termed data owners), not all of them (for instance, medical laboratory scientists) need to know patients' identity. In addition, patients' healthcare information need to be protected and the patients would define access policy specifying who have the permission to access what information.

To authenticate patients while guaranteeing anonymity, the scheme of Attribute-based Signature (ABS)~\cite{maji2008attribute} can be a potential solution. Using ABS, a signature signed with a patient's attributes is attested not to the identity of patient but instead to the attributes possessed by patient. However, ABS is not flexible to update the policy embedded in the signature since it adopts a one-time signature generation process.
% As a result, ABS enables data users to authenticate patients without revealing their identity.

To secure EHRs while facilitating versatile access and sharing among healthcare providers, the method of Attribute-based Encryption (ABE)~\cite{sahai2005fuzzy} can be a candidate solution. ABE is a public-key encryption scheme that binds security directly to EHRs and the participants who access it by enforcing attribute-based access control. Instead of encrypting EHRs for a specific data user, ABE allows encrypted EHRs to be accessed by any user with proper attributes satisfying the access policy.

Recently, the technology of Blockchain \cite{narayanan2016bitcoin} has been proposed as a solution for EHR  management~\cite{ekblaw2016case,halamka2017potential,guo2019access,griggs2018healthcare}. In this context, a blockchain can be used as a tamper-proof log to record both the authentication events on the patients and the access activities of EHRs by the data users. 
%Given the decentralized nature of a blockchain, all the events and activities would be transparent to all the participants.
However, to maximize the capability of blockchain-based EHR management solutions, the following issues need to be addressed. First, the privacy of patients and the security of their EHRs. Due to the decentralized and transparent nature of blockchain, any identity and sensitive information of the patients should not be stored directly into blockchain transactions. Second, the size of blocks in a blockchain. Typically, the size of blocks in a blockchain is too limited to accommodate EHRs containing images of X-ray, CT scans, and MRI, and videos of ultrasound.

This paper proposes the Attribute-based Multi-Signature (ABMS) scheme to authenticate patients anonymously. The paper also describes the adoption of ABE to secure EHRs with policy-based access control. Both schemes are integrated into a hybrid architecture of blockchain and edge computing to facilitate EHR management. Specifically, authentication events of patients and EHR access activities are recorded into the blockchain, i.e., {\em on-chain}, for traceability and accountability.
%Different from access policies utilized in the \cite{guo2019access}, we utilize the proposed ABMS signatures compared against the multi-signature threshold to facilitate the authentication process and access control events between patients and healthcare providers. 
In collaboration, edge nodes, which function as {\em off-chain} storage, store ABE-encrypted EHR data, so that only eligible data users satisfying the specific EHR access policies can decrypt and access the EHR data.

We prototype the designed hybrid architecture by using the Hyperledger Fabric \cite{hyperledgerfabric} and the Hyperledger Ursa \cite{hyperledgerursa}. In addition, the Access Control Lists (ACL) mechanism is used to facilitate attribute-based access control of patient profiles. Using the prototype, we conduct experiments to validate the operations of the smart contracts and policy-based access control. We also evaluate the performance of the signing and verification of multiple signatures of ABMS and blockchain transactions under different application settings.

\section{System Architecture}
\begin{figure}[ht]
\centering
\includegraphics[width=0.481\textwidth]{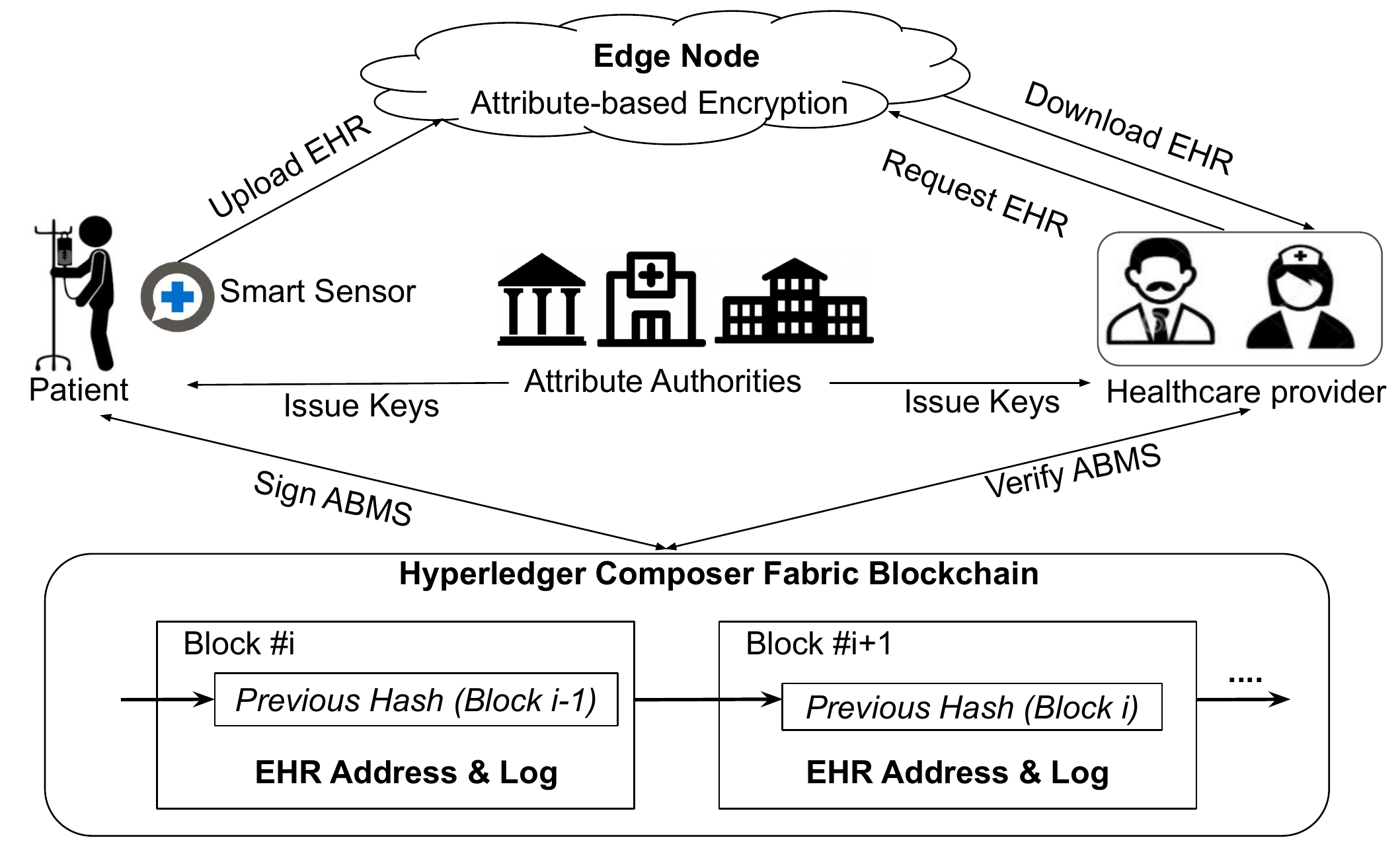}
\caption{Proposed System Architecture.}
\label{fig:arch}
\end{figure}

In this section, we describe the hybrid blockchain-edge architecture consisting of the multi-authority ABE and ABMS schemes.
%to secure patients' EHRs and preserve their identity. 
By referring to Fig. \ref{fig:arch}, we define the following entities that take part in the proposed architecture.

\begin{itemize}
\item EHR data: EHR data is a piece of information owned by a patient, and can be accessed by authorized healthcare providers who have the access permission.

\item Patient: A patient is the data owner of his/her EHR data, who defines the access policy for the data users (e.g., healthcare providers). 

\item Healthcare provider: A healthcare provider (e.g., doctor and nurse) is a data user who needs to access the EHR data owned by patients. A healthcare provider actively seeks access permissions from patients.

\item Attribute: An attribute is a piece of information (e.g., driver's license) that is associated with a participant. 

\item Attribute authority: An attribute authority is an entity who manages attributes and issues key pairs to data owners and data users.

%A patient may use his/her attributes and pre-defined access policies to encrypt the EHR data he/she owns.

%owned by patients. A healthcare provider actively verify seeks access to authorizations from patients.

\item Smart sensor: A smart sensor is a device that collects EHR data from the patients and sends it to the edge node. Smart sensors include imaging equipment such as X-ray, CT scan, MRI, and ultrasound.

\item Edge node: An edge node is a computing and storage device, which stores ABE-encrypted EHR data.  

\item Smart contracts and blockchain: Smart contracts take patients' signatures as input and return EHR addresses. The blockchain serves as a tamper-proof log of patient authentication events and EHR access activities.

%takes in charge of the attributes and provides authorization to the qualified entities.
\end{itemize}

%In the remainder of this section, we first describe the multi-authority ABE and ABMS schemes. The EHR data presented in this architecture is encrypted by the multi-authority CP-ABE scheme, whereas the patient is authenticated by the multi-authority ABMS scheme. The proposed mechanism is sufficient for the patient to prove his/her possession of attributes without revealing the identity information. In addition, we describe the on-chain and off-chain hybrid framework. Finally, we summarize the workflow of EHR management.

\subsection{Multi-authority CP-ABE mechanism design}

%add the time sensitive and dynamic model for the proposed schemes.

This subsection describes the use of multi-authority CP-ABE~\cite{lewko2011decentralizing} in the context of EHR management. 
%and the EHR information is dynamic change with the timestamp notion.

%Both data owners and data users are issued attribute keys by the attribute authorities. Each participant in EHR management obtains a global identification number $\it GID$ upon registration, which is stored in the participant's profile. The following are the steps of CP-ABE:

\textit{Initial Setup}($\lambda$) $\longrightarrow$ $\it IP$. The initial setup takes a security parameter $\lambda$ as the input and outputs the initial parameter $\it IP$ for the system.

\textit{Authority Setup}($\it IP, AT_i$) $\longrightarrow$ $PK_i$, $SK_i$. Authority $i$ takes the initial parameter $\it IP$ and attribute $AT_i$ as the input to generate the public key $PK_i$ and secret master key $SK_i$.

\textit{Key Generation}($\it AT_i, GID, {SK_i}, IP$) $\longrightarrow$ $\it {K_i, _{GID}}$. This key generation algorithm takes as inputs attribute $\it AT_i$, the global identification number $GID$ of a data user, the secret master key $SK_i$, and the initial parameter $\it IP$, and returns the attribute key $\it {K_i,_{GID}}$ to the data user.
%The private key $\it {K_i, _{GID}}$ for an attribute $\it i$ for the patient with the identity $\it GID$ is issued by including the patient's global identity  $\it GID$, the initial setup $\it IP$ and the corresponding authorities' secret master key ${SK_i}$ as the input parameters and finally generating the private key for the patient $\it i$.

\textit{Encryption}($\it IP, M, \mathbb{A}, \{{PK_i}\}$) $\longrightarrow$ $\it CT$. The encryption algorithm takes as inputs the initial parameter $\it IP$, EHR data $M$, access policy $\mathbb{A}$ defined by a patient, and the set of public keys ${PK_i}$. It outputs the ciphertext $\it CT$ as the encrypted EHR data. Moreover, the access policy $\mathbb{A}$ defines which healthcare providers are allowed to access the EHR data of the patient. 
% Whenever an EHR data of the patient is updated, the updated EHR data will be encrypted and saved as part of the patient's medical history.

\textit{Decryption} ($\it  CT, IP, \{K_i,_{GID}\}$) $\longrightarrow$ $\it M$. The decryption algorithm takes as inputs the initial parameter $\it IP$, ciphertext $CT$, and a collection of attribute keys $\{K_i,_{GID}\}$. It outputs the EHR $\it M$ if the set of attributes keys satisfies the access policy for the ciphertext $\it CT$. Otherwise, the decryption fails.

\subsection{Multi-authority ABMS mechanism design}

% In this subsection, we describe the multi-authority ABMS mechanism for  authentication. In ABMS, a patient's signatures are verified through the verification algorithm by using a multi-signature threshold scheme. The attribute-based signatures hide the identity information of the data owner (patients) with corresponding possessed attributes. We utilized the BLS signature scheme \cite{boneh2003survey} for the implementation of signing and verifying phases. 
By extending \cite{li2010attribute,shahandashti2009threshold},
the model of multi-authority ABMS is described as below:

\textit{Initial Setup}($\lambda$) $\longrightarrow$ $\it IP$: The initial setup takes the security parameter $\lambda$ as input and outputs the initial parameters $\it IP$ for the system.
%At the same time, this algorithm generates a secret key for each of the registered attribute authorities, and also a master secret key for the central authority.

\textit{Authority Setup}($\it IP, AT_i$) $\longrightarrow$ $\it {SIK_i}, {VK_i}$: Each attribute authority executes the setup algorithm with the input of the initial parameter $\it IP$ and attribute $AT_i$, and returns the signature key $\it SIK_i$ and the verification key $\it VK_i$ for each attribute $\it AT_i$.

\textit{Extract}($\it IP, GID, AT_i, {SIK_i}$) $\longrightarrow$ $\it {SK_i, _{GID}}$: This algorithm is executed by the attribute authorities. An attribute authority takes as inputs the initial parameters $\it IP$, the owner's unique identity $\it GID$, the attribute $AT_i$, and the authority's signature key ${SIK_i}$. In the end, the algorithm returns the signing key $\it {SK_i,_{GID}}$.

\textit{Sign}($\it H(A_i), IP, SK_i,_{GID}$) $\longrightarrow$ $\sigma_i$: This algorithm is executed $\it n$ times based on the number of attributes belonging to the data owner (patients). It takes as inputs the hashed attribute value $\it H(A_i)$, initial parameter $\it IP$, and signing key $\it {SK_i,_{GID}}$, and outputs the signature $\sigma_i$.
%The data owner signs the attributes information $\it A^{\tau}$, and takes the secret key for these attributes where predicate $\Upsilon = 1$, then outputs the signature $\sigma^{\tau}_i$.
%Note that the notion of $\it \tau$ captures the proposed time-sensitive and dynamic nature of our ABMS scheme.

\textit{Verify}($\it H(A_i), IP, \sigma_i, VK_i$) $\longrightarrow$ $\{0,1\}$: This algorithm takes as inputs the hashed attribute value $\it H(A_i)$, the initial parameter $\it IP$, the signature $\sigma_i$, and the verification key $VK_i$.
In the end, the algorithm outputs a Boolean value $\it accept$ or $\it reject$ to indicate whether the signature from the data singer (patient) with the specific attribute is valid or not without revealing the signer's identity.

%In the following, we describe the signing and verifying algorithms in more details, and we show the proof of the correctness of the proposed scheme. The ABMS signing process: ($\it M^{\tau}, IP, \Upsilon, \{{K_i,_{GID}\}}$) $\longrightarrow$ $\sigma^{\tau}_i$. The signature $\sigma^{\tau}_i$ for the encrypted EHR message $\it M^{\tau}$ can be generated when the corresponding set of attributes $A$ satisfies the predicate $\Upsilon(A) = 1$ and the correct $\{{K_i,_{GID}}\}$ key information which is generated by the key generation algorithm. Next, the ABS signature $\sigma^{\tau}_i$ is verified to be $\it True$ when the output of the ABMS verification function equals to 1 by the verification algorithm otherwise it will return the false information. Interested readers can refer to \cite{li2019blockchain}, \cite{cao2011multi}, \cite{okamoto2013decentralized} for more detailed information and formal mathematical proof of the correctness of the signing and verification algorithms.

%Shamir secret sharing mechanism and $\it (t,n)$ threshold signature scheme. In real-world scenario, the $\it t$ is in the range of $\it (1, n)$. 

\begin{figure}[ht]
\centering
\includegraphics[width=0.47\textwidth]{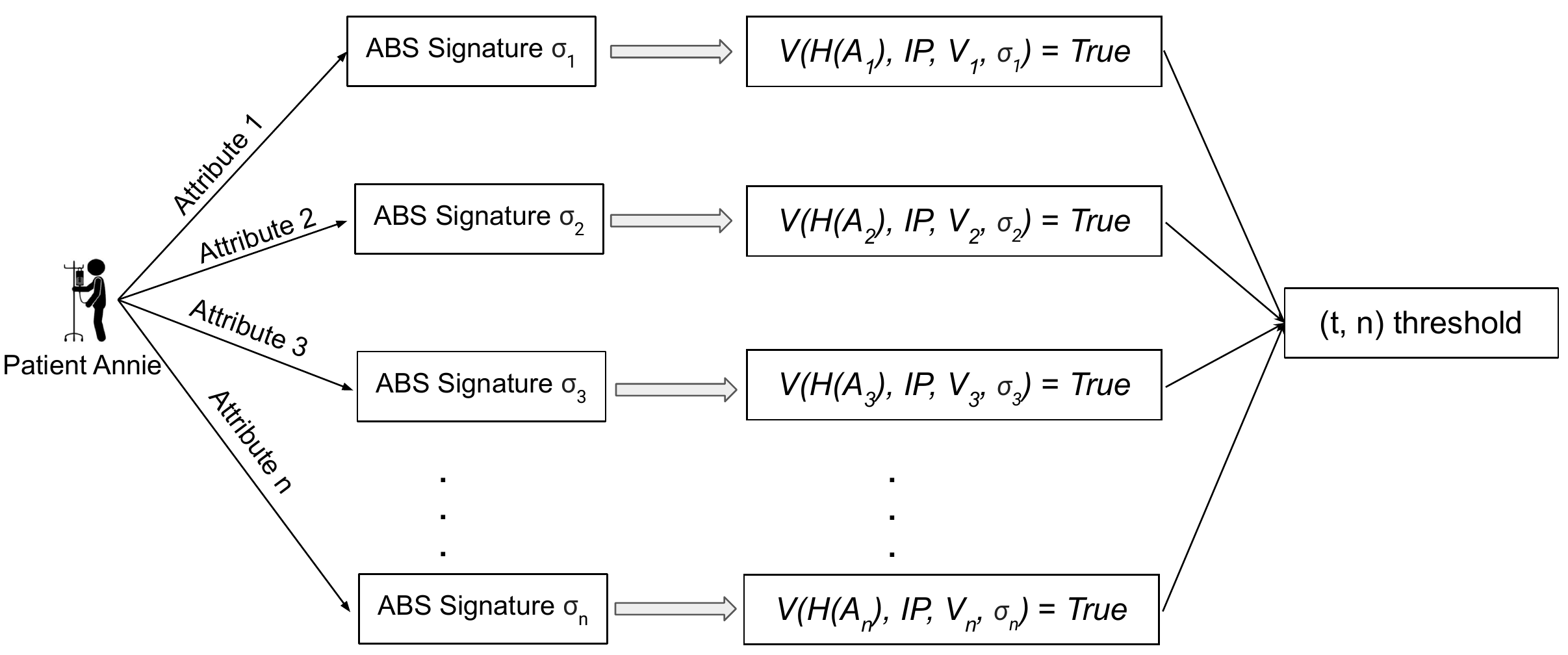}
\caption{Multi-signature threshold scheme.}
\label{fig:multisig}
\end{figure}

In the proposed ABMS scheme, we applied the $\it (t, n)$ threshold scheme, where $1\leq t \leq n$, to as shown in Fig.~\ref{fig:multisig}. 
For instance, patient Annie signed three signatures related to her patient ID issued by the hospital, her driver's license issued by DMV, and her insurance ID issued by the insurance company. A medical laboratory scientist, who does test on Annie's blood work, will apply the threshold of {\em 3 out of 3} to authenticate her, while a medical research scientist will apply the threshold of {\em 1 out of 3} to authenticate Annie to access her EHR for data analysis. 
% Next, healthcare provider can verify patient's signatures against the threshold. If the threshold value $\it t$ has been increased, the proposed multi-signature scheme requires higher number of signatures to be authenticated by the healthcare provider. 
For both cases, the aim of ABMS is to authenticate patients anonymously while using different thresholds. 

%we utilize the Shamir's secret sharing~\cite{shamir1979share} mechanism to secure the signature information in a distributed way. The signatures are split into multiple parts, called shares. These shares are used to reconstruct the original signature information. The aim is to divide the signature $\sigma$ into $\it n$ pieces of the signatures $\sigma_{1}, \sigma_{2}, ... , \sigma_{n}$ in the way that:
%1. The knowledge of any $\it t$ or more $\it \sigma_{i}$ pieces of signature information make $\sigma$ feasible to compute and indicates that the complete signature can be reconstructed from any combination of $\it t$ pieces of the signature information. 2. The knowledge of any $\it t-1$ or even less pieces of signature $\it \sigma_{i}$ will make $\sigma$ to be undermined which indicate that the signature $\sigma$ cannot be reconstructed with fewer than $\it t$ pieces of the signature information. The scheme is named $\it (t, n)$ threshold secret sharing mechanism. For instance, if $\it t = n$ then every piece of signature $\sigma_{i}$ is needed to satisfy the condition to reconstruct the secret. 

%\subsection{Blockchain Network Model}

%In this subsection, we introduce the Hyperledger Fabric blockchain network architecture for our proposed mechanism, and introduce the smart contracts with different data structure and consensus algorithm. To implement and evaluate the multi-authority ABS and ABE mechanism, we designed and developed a web-based blockchain application on the Hyperledger  Fabric network. 

\subsection{On-chain and off-chain hybrid architecture}

\begin{figure}[ht]
\centering
\includegraphics[width=0.46\textwidth]{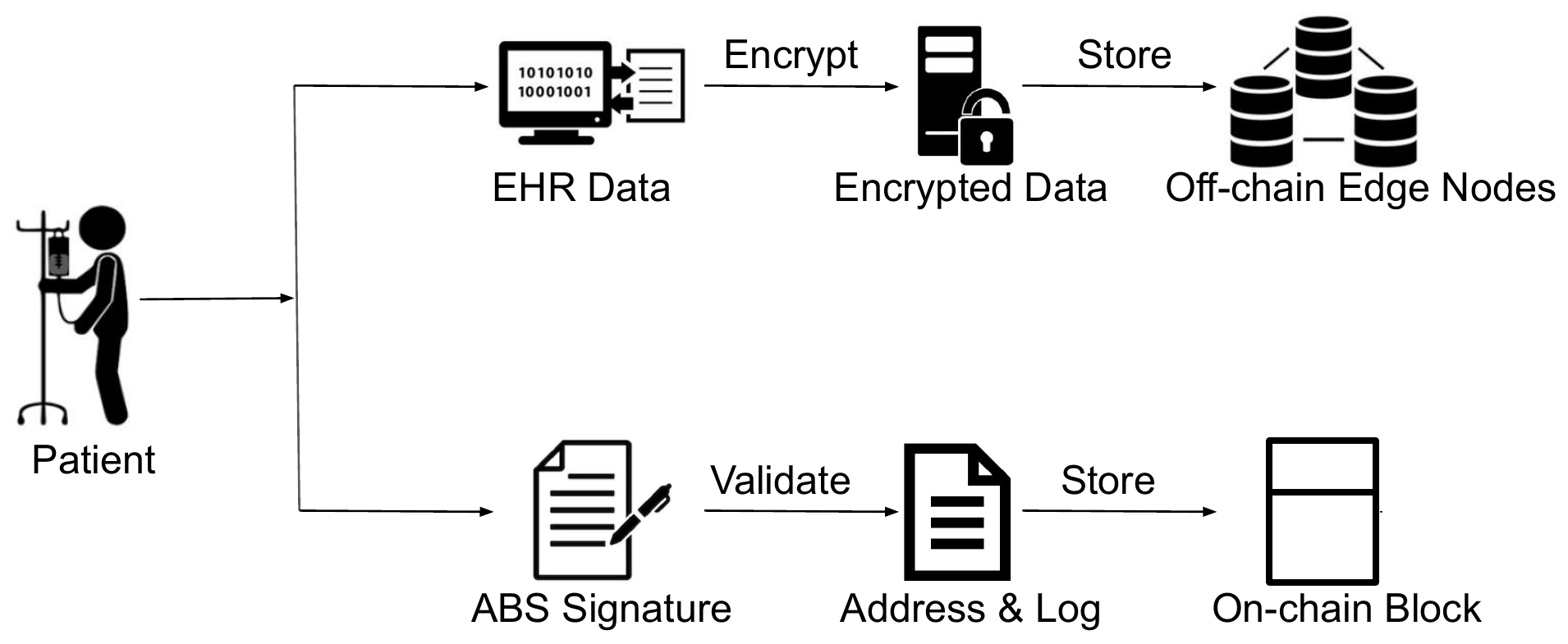}
\caption{Hybrid on-chain and off-chain architecture.}
\label{fig:onoff}
\end{figure}

To overcome the space limitation issue~\cite{guo2019access}, the transaction privacy issues~\cite{zyskind2015decentralizing,baza2019b,baza2018blockchain}, and the issue of lacking proper access control among participants~\cite{guo2019access,guo2019multi} of the blockchain, we propose a hybrid architecture which consists of an on-chain record of ABMS authentication events and EHR access activities (including EHR addresses and other information) as transactions on a blockchain, and an off-chain storage of ABE-encrypted EHR data on the edge nodes, as shown in Fig. \ref{fig:onoff}. 
%Blockchain transactions are transparent to every participants where we only store the EHR addresses and access log.

In addition, each patient saves his/her data, including $\it GID$, name, and ABMS signatures in a private profile. 
%the patient can sign the time-sensitive EHR data with the private key which is related to her attributes and stores the ABE-encrypted EHR data on off-chain edge node.
A patient will define access policy for the data users (e.g., healthcare providers) to access his/her profile, which is implemented with the ACL mechanism of the Hyperledger Fabric blockchain.
%generating the address of the encrypted EHR data with her attributes and publish the digital signature information into smart contract which is stored on the blockchain.

\subsection{Workflow of EHR management}

\begin{figure}[ht]
\centering
\includegraphics[width=0.485\textwidth]{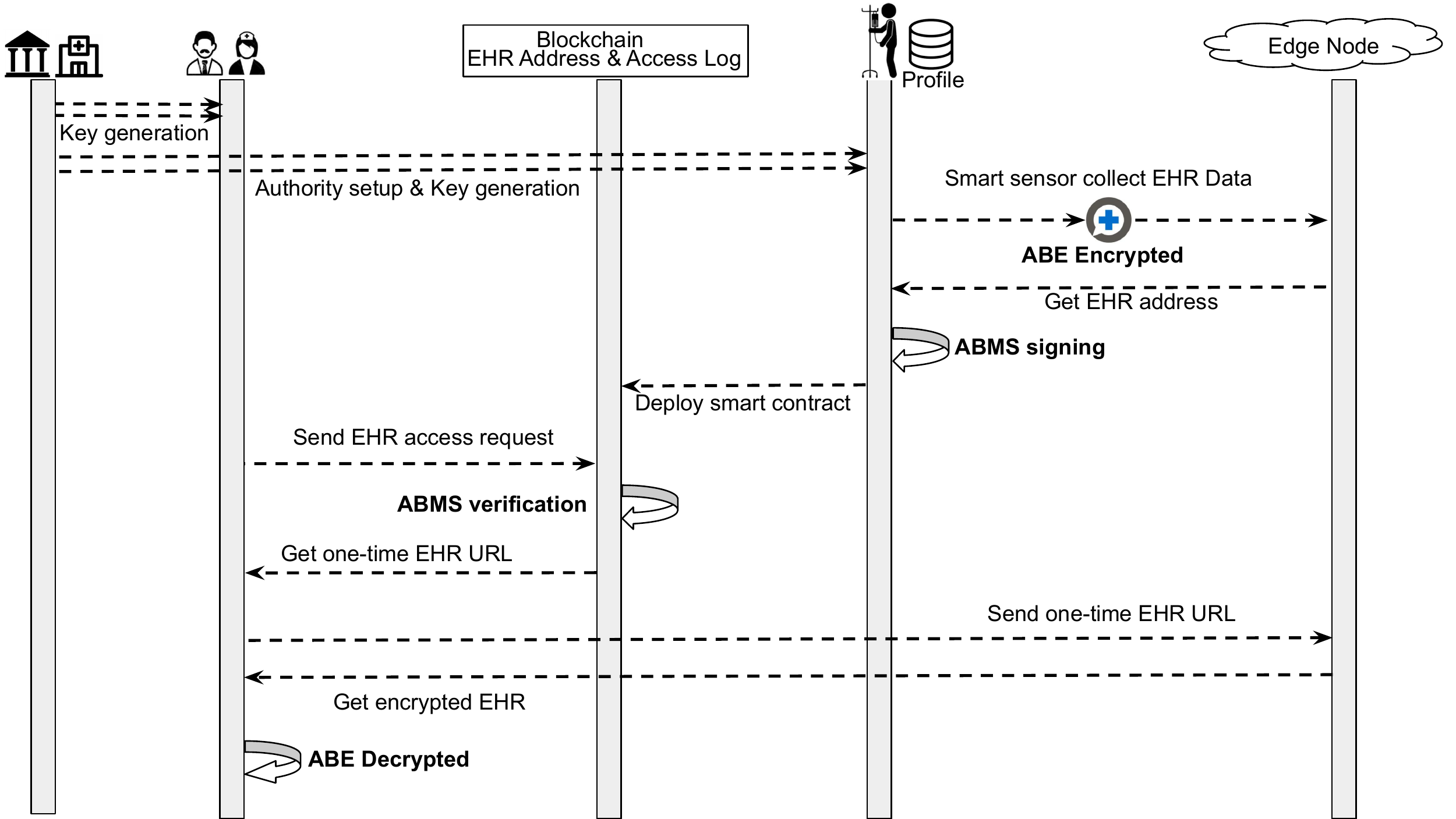}
\caption{Workflow of Attribute-based Multi-Signature and Encryption for EHR management.}
\label{fig:workflow}
\end{figure}

Workflow of EHR management is depicted in Fig. \ref{fig:workflow}. First, all the participants register themselves with the EHR management system. Attribute authorities issue ABMS and ABE keys to both patients and healthcare providers based on their attributes. After being granted private keys from multiple attribute authorities, a patient uploads his/her EHR data, which is encrypted via ABE\footnote{In practice, it is not practical for patients to handle (e.g., encrypt) their EHR data personally. To address this issue, smart sensors could be the entities to receive these private keys to ABE-encrypt EHR data before uploading it to the edge node.}, to the edge node. Next, a patient signs his/her hashed attribute values to generate the ABMS signatures, which are stored in his/her profile in the EHR management system.  When a data user (e.g., a healthcare provider) needs to access a patient's EHR data, an EHR access request is sent to the blockchain which triggers ABMS signature verification to authenticate the patient. A smart contract is executed to log the authentication event and return a one-time EHR URL to the healthcare provider.  After that, the healthcare provider sends the one-time EHR URL to the edge node to retrieve the encrypted EHR data.
%As long as the healthcare provider possesses the attributes required to satisfy the access policies specified by the patient, he/she can decrypt the EHR data received from the edge node.

\section{Prototype and Evaluation}

\subsection{Prototype}

We prototyped the proposed EHR management system and conducted a series of experiments to evaluate its performance. The system consists of two primary modules that interact seamlessly: the ABMS module and the blockchain module.
%The ABMS module was programmed by using the Hyperledger Ursa library \cite{hyperledgerursa} and deployed and experimented on a desktop with a 2.8 GHz Intel i5-8400 processor and 8GB of memory running Ubuntu 18.04. The blockchain module was developed on the Hyperledger Fabric platform \cite{hyperledgerfabric} and ran on the Hyperledger Composer Playground.

\subsection*{A.1 ABMS Module}

%As depicted in Fig. \ref{fig:workflow},
ABMS performs the functionalities of key generation, and signing and verification of attribute-based multi-signatures of the patients. These functionalities are programmed by using Hyperledger Ursa \cite{hyperledgerursa}, a cryptographic library for blockchain-based applications. Hyperledger Ursa is built on the Rust language and provides APIs for various digital signature schemes.

%The ABMS module operates in the following four phases as shown in Fig. \ref{fig:ABMS-process}. 

% including instantiating, generating key pairs, signing, and verifying ABMS.

\subsubsection{Phase 1 -- Initialization}
Phase 1 initializes the instances of the participants which include multiple authorities, patients, and doctors.
%As shown in Fig. \ref{fig:ABMS-process},
The patient Annie Foster, for instance, has three attributes, which are {\tt patient\_id} (value: {\tt 0003231}) issued by the hospital, {\tt driver\_license} (value: {\tt 9907184}) issued by the Department of Motor Vehicles (DMV), and {\tt insurance\_id} (value: {\tt 1EG4-TE5-MK72}) issued by her health insurance company.

%\begin{figure}[ht]
%\centering
%\includegraphics[width=0.488\textwidth]{Figures/ma-abms-process-5.png}
%\caption{Process of the ABMS module.}
%\label{fig:ABMS-process}
%\end{figure}

\subsubsection{Phase 2 -- Multi-Authority Key Generation}
In Phase 2, each attribute authority generates a key pair for each attribute it issues. The BLS signature scheme \cite{boneh2001short} was used to build the key-pair generator, which generates the signing key for the data owner and the verification key for the data user, as follows:

%12-13: I double checked the source code. The following code is correct.

\begin{verbatim}
let generator = Generator::new().unwrap();
let sign_key = SignKey::new().unwrap();
let verify_key = VerKey::new(&generator,
                 &sign_key).unwrap();
\end{verbatim}

%The BLS scheme uses a bilinear pairing for verification, and signatures are elements of an elliptic curve group. 

\subsubsection{Phase 3 -- Signing ABMS Signatures by Data Owner}
\begin{figure}[h]
\centering
\includegraphics[width=0.39\textwidth]{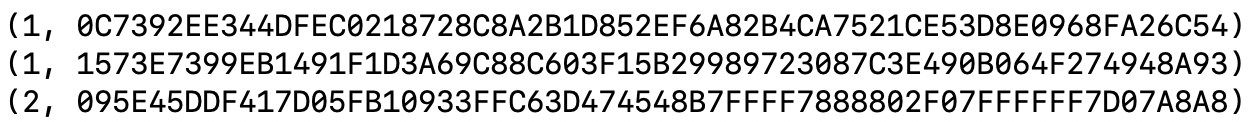}
\caption{Example of one ABMS signature.}
\label{fig:signature-example}
\end{figure}

In this phase, the data owners use the signing keys to sign their hashed attribute values, and the resulting signatures are saved in data owners' profiles on the blockchain (Section IV-A.2). Using the BLS signature scheme \cite{boneh2001short}, each signature consists of three elements on an elliptic curve. For instance, as shown in Fig. \ref{fig:signature-example}, the hashed attribute value of {\tt 0003231} from {\tt patient\_id} was signed into a combination of three points on an elliptic curve. Our experiments show that the average running time for the signing phase is around 32 ms.

\subsubsection{Phase 4 -- Verifying ABMS Signatures by Data User}
\label{sec:phase4}

Eventually, a data user verifies each ABMS signature retrieved from the blockchain (Section IV-A.2). The verification function takes a signature, a hashed attribute value, a verification key and the corresponding generator as inputs, and utilizes BLS bilinear pairing \cite{boneh2003survey} to verify the signature:

%12-13: I double checked the source code. The following code is correct.

\begin{verbatim}
let result = Bls::verify($signature, 
             patient_id.as_slice(),
             $verify_key, $generator)
             .unwrap();
\end{verbatim}

In our experiments, the average running time for verifying each signature is around 243 ms. After that, the data user can authenticate the data owner anonymously by applying the multi-signature threshold mechanism, where the number of verified signatures is compared against the threshold. 

\subsection*{A.2 Blockchain Module}
\label{sec:blockchain_module}

\begin{comment}
\begin{figure}[h]
\centering
\includegraphics[width=0.46\textwidth]{Figures/login-window.png}
\caption{Blockchain-based EHR network login window.}
\label{fig:login-window}
\end{figure}
\end{comment}

The blockchain module is developed on the Hyperledger Fabric and deployed and executed on the Hyperledger Composer Playground \cite{hyperledgerfabric}. The blockchain module records patients' profiles of GIDs, first and last names, signed ABMS signatures, and one-time self-destructing url addresses \cite{1ty} for EHR data stored on the edge node. We use {\tt https://1ty.me/}~\cite{1ty} to encode the addresses of EHR data stored on the edge node. Once an {\tt 1ty.me} address is accessed, its url link becomes invalid and cannot be used to access the same EHR data again.  

%\begin{figure}[h]
%\centering
%\includegraphics[width=0.488\textwidth]{Figures/doctor-access-2.png}
%\caption{A doctor accesses a list of patient profiles.}
%\label{fig:doctor-access}
%\end{figure}

% Fig. \ref{fig:login-window} shows an example with one doctor and five patients as participants in blockchain network.

%where each participant has a unique digital ID card to log into the blockchain network. 

\begin{figure}[h]
\centering
\includegraphics[width=0.38\textwidth]{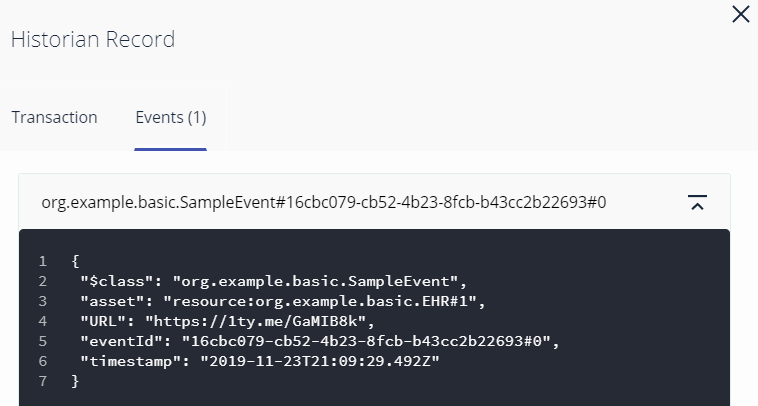}
\caption{Result of executing the smart contract.}
\label{fig:transaction-detail}
\end{figure}

To evaluate the access control mechanism between patients and healthcare providers, we conducted the following experiments.
%A doctor is allowed to view all of his patients' profiles, including GIDs, first names, last names and signed ABMS signatures as shown in Fig. \ref{fig:doctor-access} by satisfying the acccess polices defined in the ACL. However, a patient is restricted to access his/her own profile.
When a healthcare provider wants to access a specific patient’s EHR data from the off-chain storage, he/she needs to first authenticate the patient by verifying patient's ABMS signatures (Section IV-A.1(4)) against the threshold via a smart contract. Upon executing the smart contract successfully, it returns a one-time self-destructing url address of patient’s EHR data stored in the edge node. At the same time, the blockchain will  record this access event as a transaction, which includes the event ID and the timestamp as shown in Fig. \ref{fig:transaction-detail}.
%Therefore, any EHR access event can be traced back for further analysis.

\subsection{Performance Evaluation}

To evaluate the performance of the ABMS module, we conducted experiments to analyze the effect of varying the length of attribute values as well as the number of attributes. 
% In addition, we evaluated the blockchain module by recording the transaction processing time and the response time against unauthorized retrievals under different settings.

%and three aspects of the two modules. For the ABMS scheme module, we tested the running time of signing and verifying phases by changing the length and the number of attributes. For the Blockchain network module

\subsubsection{Varying the length of attributes}

%\begin{figure}[h]
%\centering
%\includegraphics[width=0.489\textwidth]{Figures/attribute-length-2.png}
%\caption{Run time of Ursa BLS vs. the length of attributes.}
%\label{fig:attribute-length}
%\end{figure}

We increased the length of attribute from 10 to 100, 1,000, and 10,000 characters, and the results 
%(Fig. \ref{fig:attribute-length})
showed that both signing and verification time are independent of the attribute length, and the time remains around 32 ms and 243 ms, respectively, for each attribute. 
% This is due to the fact that the Hyperledger Ursa BLS scheme generates fixed-length signatures as shown in Fig. \ref{fig:signature-example}. 

\subsubsection{Varying the number of attributes}

\begin{figure}[h]
\centering
\includegraphics[width=0.4\textwidth]{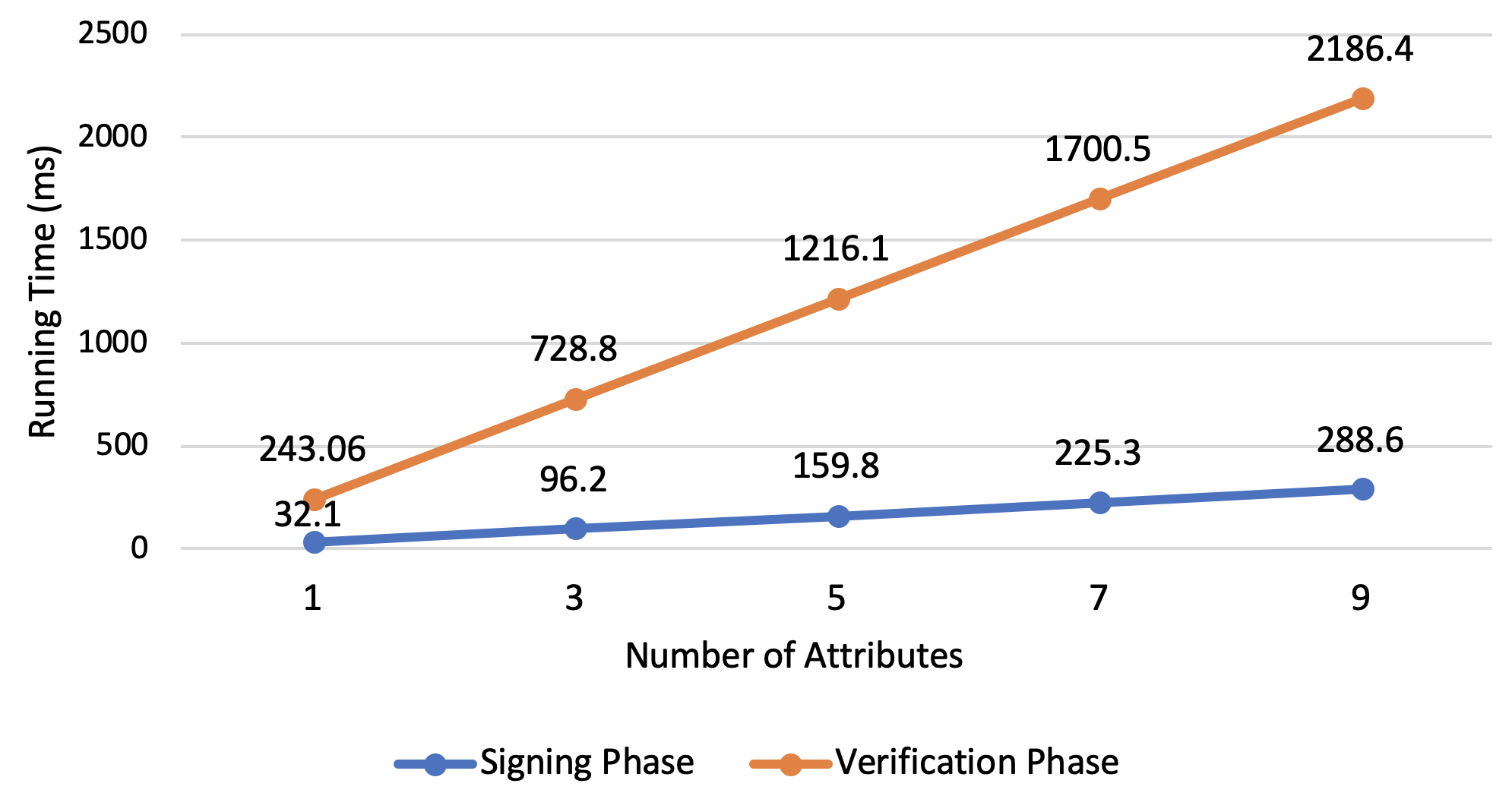}
\caption{Run time of Ursa BLS vs. the number of attributes.}
\label{fig:attribute-number}
\end{figure}

We also increased the number of attributes of a patient from 1 to 3, 5, 7 and 9, and measured the running time of the signature signing and verification phases. The total running time showed a linear growth with the increasing the number of attributes (Fig. \ref{fig:attribute-number}). Compared to the signing phase, the verification phase takes more time because it requires computing two pairings on the elliptic curve \cite{boneh2001short}.

\section{Summary}
\label{conclusion}

%The emerging eHealth smart sensors technologies (e.g., health monitoring wearable devices) enable EHR systems to collect and communicate health data via different platforms. While this is significantly beneficial for patients' health analysis and monitoring, it raises new privacy and security concerns motivating the need for designing and developing fine-grained access control systems for EHR data management.

In this paper, we proposed a hybrid architecture of blockchain and edge nodes, by utilizing the ABMS scheme to authenticate user's signatures without revealing sensitive information
%achieve user's authenticity without revealing sensitive identity information 
and the ABE mechanism to encrypt EHR data which is stored on the edge node.
%The architecture utilizes blockchain to (1) execute smart contracts to impose the ABMS mechanism, and (2) record EHR data storage addresses and access logs permanently. 
%In addition, EHR data are stored on edge nodes, which are protected by the ABE mechanism.
We developed the blockchain module on Hyperledger Fabric platform and the ABMS module on Hyperledger Ursa library. To evaluate the system performance, we designed and conducted experiments for ABMS module. For ABMS module, we measured the signing and verifying time under different settings. For the blockchain module,
we experimented with the authentication events and access activities, which were logged as transactions in the Hyperledger Fabric blockchain.
%we measured the performance of blockchain transactions under both authorized and unauthorized access attempts.

%The experimental results show that our system can response in terms of milliseconds, making it suitable to be incorporated in real-time and secured EHR data access control frameworks.
% For future work, we plan to investigate novel consensus protocol designs for the proposed system to achieve better performance. 
%In addition, we plan to develop a Hyperledger-based benchmark tool for a set of performance analyses.

% conference papers do not normally have an appendix

% use section* for acknowledgment

% trigger a \newpage just before the given reference
% number - used to balance the columns on the last page
% adjust value as needed - may need to be readjusted if
% the document is modified later
%\IEEEtriggeratref{8}
% The "triggered" command can be changed if desired:
%\IEEEtriggercmd{\enlargethispage{-5in}}

% references section

% can use a bibliography generated by BibTeX as a .bbl file
% BibTeX documentation can be easily obtained at:
% http://www.ctan.org/tex-archive/biblio/bibtex/contrib/doc/
% The IEEEtran BibTeX style support page is at:
% http://www.michaelshell.org/tex/ieeetran/bibtex/
%\bibliographystyle{IEEEtran}
% argument is your BibTeX string definitions and bibliography database(s)
%\bibliography{IEEEabrv,../bib/paper}
%
% <OR> manually copy in the resultant .bbl file
% set second argument of \begin to the number of references
% (used to reserve space for the reference number labels box)

\bibliographystyle{IEEEtran}
\bibliography{bibtex/IEEEabrv,bibtex/sig}

% that's all folks
\end{document}